\documentclass[preprint,prd,amsmath,amssymb,nofootinbib,tightenlines,floatfix,letterpaper]{revtex4}
\usepackage[dvips]{graphics}
\usepackage{epsfig}
\usepackage{latexsym}
\newcommand{\be}{\begin{equation}}
\newcommand{\ee}{\end{equation}}
\newcommand{\bea}{\begin{eqnarray}}
\newcommand{\eea}{\end{eqnarray}}

\begin{document}
\preprint{BNL-HET-08/18}

\title{Model-Independent Constraints on Lepton-Flavor-Violating Decays of the Top Quark}

\author{Jennifer Kile\footnote{
	Electronic address: jenkile@quark.phy.bnl.gov}
}
\affiliation{Brookhaven National Laboratory, Upton, NY 11973}

\author{Amarjit Soni\footnote{
	Electronic address: soni@bnl.gov}}
\affiliation{Brookhaven National Laboratory, Upton, NY 11973}

\begin{abstract}
The imminent start of the Large Hadron Collider, which is expected to produce $\sim 10^8$ $t\bar{t}$ pairs per year, provides an unprecedented opportunity for top physics.  As the top quark is widely expected to be rather sensitive to effects of new physics, a detailed study of its properties, including rare decays, is called for.  A possible, experimentally distinctive decay is the case where a top decays to a light quark and a flavor-violating lepton-antilepton pair.  We use an effective operator analysis to place model-independent bounds on contributions to the decays $t\rightarrow u e^{\pm} \mu^{\mp}$ and $t\rightarrow c e^{\pm} \mu^{\mp}$.  We enumerate the dimension-six operators which contribute to these decays and which are invariant under the Standard Model gauge group.  We separate these operators into two classes, one with operators where the top quark belongs to an $SU(2)$ doublet and thus can contribute at tree level to low-energy processes, and one class with operators where the top quark is a right-handed singlet and can only contribute to low-energy processes via loop diagrams.  We use $B$ and $K$ decays to place limits on the coefficients of some of these operators, but find that several remain unconstrained and could potentially make observable contributions to top decay.

\end{abstract}

\maketitle

\section{Introduction}
\label{sec:intro}

The startup of the Large Hadron Collider will give an unprecedented opportunity to improve our knowledge of physics at energies above the weak scale.  Given the recent measurements that show that two of the neutrino mixing angles are large \cite{Ashie:2004mr,Aharmim:2005gt,Abe:2008ee}, it may be useful to consider the possibility that new physics (NP) could display deviations from the lepton flavor conservation and the approximate quark flavor conservation of the Standard Model (SM).  One manifestation of such NP, possibly observable at LHC, is flavor-violating decays of the top quark.  As we wish to make no assumptions about the flavor structure of the physics beyond the Standard Model, we choose a specific decay to study based on the distinctiveness of its experimental signature.  Thus, we consider the decay of a top quark to a light, up-type quark, an electron, and an antimuon (or vice versa), $t\rightarrow u(c) e^{\pm} \mu^{\mp}$.  

Many previous studies have considered the experimentally similar, lepton-flavor-conserving processes  $t\rightarrow u(c) Z$ and $t\rightarrow u(c) \ell^+ \ell^-$ in the SM, MSSM, two-Higgs-doublet models, and in many other extensions \cite{Eilam:1990zc,Frank:2006ku}. (See also references in  \cite{Frank:2006ku}; for model-independent analyses, see \cite{Fox:2007in,AguilarSaavedra:2004wm,Ferreira:2008cj,Han:1995pk}.)  Top decays which violate lepton number, meanwhile, have been studied within the context of two-Higgs-doublet models \cite{Iltan:2002re,Iltan:2004js,DiazCruz:2004sp}, topcolor models \cite{Yue:2004jc}, and  the MSSM, models with flavon-Higgs mixing, and strongly interacting models  \cite{DiazCruz:2004sp}.  However, in an effort to be as model-independent as possible, we study this decay through an effective operator analysis.  Such analyses have been done previously for top decays \cite{Gounaris:1996vn,Fox:2007in,Ferreira:2008cj} (including lepton-flavor-violating decays \cite{Pagliarone:2008af}) as well as $t\bar{t}$ production at LHC \cite{Ferreira:2006in} and single $t$ production at LHC \cite{Ferreira:2005dr,Ferreira:2006xe,Ferreira:2006jt,Ferreira:2006in,AguilarSaavedra:2008gt,Cao:2007ea}.  We assume that there exists some NP, originating at a high energy scale $\Lambda$, which can lead to $t\rightarrow u(c) e^{\pm} \mu^{\mp}$.  While we do not assume any particular form which the NP will take, we expect it to show up as some set of effective operators ${\cal O}_i$, as long as $\Lambda>>m_t$.  Each of these operators carries a coefficient $C_i$ and each is suppressed by a factor of $\Lambda^{d-4}$, where $d$ is the dimension of the operator:
\be
{\cal L}_{eff} = \sum_i \frac{C_i}{\Lambda^{d-4}} {\cal O}_i + h.c.
\ee
As operators with higher dimensions are suppressed by increasing powers of $\Lambda$, we restrict ourselves to operators with the lowest dimension relevant to the $t\rightarrow u(c) e^{\pm} \mu^{\mp}$ process, $d=6$.  Once we have a list of the operators which can contribute to $t\rightarrow u(c) e^{\pm} \mu^{\mp}$, we consider their contributions to low-energy processes.  Those operators which are constrained by low-energy processes receive their tightest limits from $B$ and $K$ decays, although we also consider the processes $\mu\rightarrow e$ conversion in nuclei, and $\mu\rightarrow e \gamma$.  These constraints allow us to place limits on the operator coefficients $C_i$, and, in turn, on the contributions these operators can make to $t$ decay.  We find that, although some operators are constrained to contribute negligibly to $t$ decay, there do remain several operators which could have sizeable effects.

We organize the rest of this paper as follows.  In Section \ref{sec:ops}, we enumerate our operator basis, separating the operators which can contribute to low-energy processes at tree level from those which cannot.  In Section \ref{sec:tops}, we discuss the reach of top decays at LHC in constraining or measuring the coefficients of these operators.  In Section \ref{sec:bs}, we place limits on some of these operators via their tree-level contributions to the decays of $B$ mesons, and in Section \ref{sec:ks} we similarly derive limits from tree-level and one-loop contributions to the decay of the $K_L$.  In Section \ref{sec:muegamma}, we briefly discuss possible constraints on operators arising from limits on the processes $\mu\rightarrow e \gamma$ and $\mu\rightarrow e$ conversion in nuclei.  Finally, in Section \ref{sec:conc}, we give our conclusions. 

\section{Operator Basis}
\label{sec:ops}
We enumerate all possible dimension-six effective operators which contribute to the process $t\rightarrow u(c) e^{\pm} \mu^{\mp}$ and which are invariant under the $SU(3) \times SU(2) \times U(1)$ gauge group of the Standard Model.  For a complete list of dimension-six $SU(3) \times SU(2) \times U(1)$-invariant operators, see \cite{Buchmuller:1985jz}.  First, we list those which we denote as the {\it Class One} operators, which contain right-handed top quarks:
\begin{eqnarray}
{\cal O}_{1,ijk} &=& \bar{u}_R^i \gamma^{\mu} t_R \bar{L}^j_L \gamma_{\mu} L^k_L \nonumber\\
{\cal O}_{2,ijk} &=& \bar{u}_R^i \gamma^{\mu} t_R \bar{l}^j_R \gamma_{\mu} l^k_R \nonumber\\
{\cal O}_{3,ijk} &=& \epsilon^{ab} \bar{Q}_{La}^i  t_R \bar{L}^j_{Lb}  l^k_R \nonumber\\
{\cal O}_{4,ijk} &=& \epsilon^{ab} \bar{Q}_{La}^i \sigma^{\mu\nu} t_R \bar{L}^j_{Lb} \sigma_{\mu\nu} l^k_R \nonumber
\end{eqnarray}
where $i$, $j$ and $k$ are flavor indices, $a$ and $b$ are $SU(2)$ indices, and $Q_L$ and $L_L$ denote the left-handed quark and lepton $SU(2)$ doublets, while $t_R$, $u_R$ and $l_R$ denote the right-handed singlet top quark, light up-type quark and charged lepton, respectively.  $i$ can stand for either of the first two families of quarks, and $j$, $k$ take on the values $e$, $\mu$ or $\mu$, $e$.  Due to the presence of the right-handed top quark, these operators do not contribute at tree level to low-energy processes.  These contrast with the {\it Class Two} operators
\begin{eqnarray}
{\cal O}_{5,ijk} &=& \bar{Q}^i_L \gamma^{\mu} T_L \bar{L}^j_L \gamma_{\mu} L^k_L \nonumber\\
{\cal O}_{6,ijk} &=& \bar{Q}^i_L \gamma^{\mu} T_L \bar{l}^j_R \gamma_{\mu} l^k_R \nonumber\\
{\cal O}_{7,ijk} &=& \epsilon^{ab} \bar{u}_R^i T_{La} \bar{l}^j_R  L^k_{Lb} \nonumber\\
{\cal O}_{8,ijk} &=& \epsilon^{ab} \bar{u}_R^i \sigma^{\mu\nu} T_{La} \bar{l}^j_R \sigma_{\mu\nu} L^k_{Lb} \nonumber
\end{eqnarray}
which contain the third-family quark doublet (denoted here $T_L$).  We explicitly note that the $SU(2)$ doublets in these operators contain the same fields as those which participate in the weak interaction; for example, $T_L$ has as its upper component, $t_L$, and, as its lower component, $V_{td} d_L + V_{ts} s_L + V_{tb} b_L$, where $t$, $d$, $s$, and $b$ are mass eigenstates.  Thus, the {\it Class Two} operators contribute at tree level to processes containing $b$ quarks and (with CKM-suppressed amplitudes) to processes containing $d$ and $s$ quarks.

\section{Top Decays}
\label{sec:tops}
\begin{figure}[h]
\epsfxsize=2in
\epsfig{figure=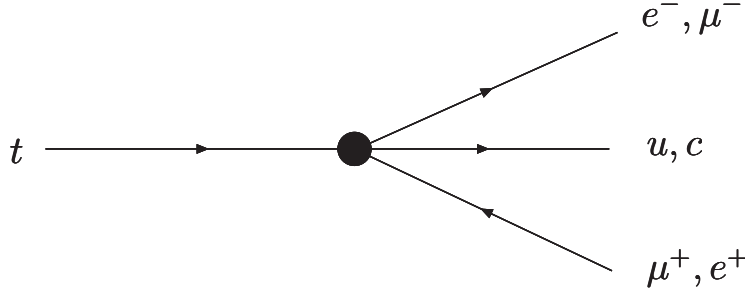,width=4.in}
\caption{Contribution of any of the dimension-six operators to $t\rightarrow u(c) e^{\pm} \mu^{\mp}$.}
\label{fig:top}
\end{figure}

We begin by considering the contributions that each of our operators can make to the process $t\rightarrow u(c) e^{\pm} \mu^{\mp}$.  As LHC is expected to produce on the order of $10^8$ top quark pairs, we calculate the values of $|C_{n,ijk}|/\Lambda^2$ such that the operators would give a branching fraction for $t\rightarrow u(c) e^{\pm} \mu^{\mp}$  of $10^{-7}$. 
The tree-level contributions of these operators (shown in Fig.\ref{fig:top}) to $t\rightarrow u(c) e^{\pm} \mu^{\mp}$ are given as follows:
\begin{eqnarray}
&& \frac{|C_{n,ijk}|^2}{\Lambda^4} \frac{1}{1536 \pi^3} m_t^5 \quad\mbox{(for $n=$ $1$, $2$, $5$, and $6$)}\nonumber\\
\Gamma(t\rightarrow u(c) e^{\pm} \mu^{\mp}) = && \frac{|C_{n,ijk}|^2}{\Lambda^4} \frac{1}{6144 \pi^3} m_t^5 \quad\mbox{($n=$ $3$ and $7$)}\label{eq:ts}\\
&& \frac{|C_{n,ijk}|^2}{\Lambda^4} \frac{1}{128 \pi^3} m_t^5 \quad\mbox{($n=$ $4$ and $8$)}\nonumber
\end{eqnarray}
where the results are independent of the flavor indices $i$, $j$, and $k$.  If we set $|C_{n,ijk}|=1$, we obtain, for the NP scale,
\begin{eqnarray}
&& 2.1 \quad\mbox{TeV}\quad\mbox{($n=$ $1$, $2$, $5$, and $6$)}\nonumber\\
\Lambda \ge && 1.5 \quad\mbox{TeV}\quad\mbox{($n=$ $3$ and $7$)}\nonumber\\
&& 4.0 \quad\mbox{TeV}\quad\mbox{($n=$ $4$ and $8$)}\nonumber
\end{eqnarray}
where we have taken $m_t = 170$ GeV.

We now wish to compare these expected constraints from top decay with constraints which we can obtain from existing limits on NP contributions to low-energy processes.  We start with decays of $B$ mesons. 

\section{B Decays}
\label{sec:bs}
\begin{figure}[h]
\epsfxsize=2in
\epsfig{figure=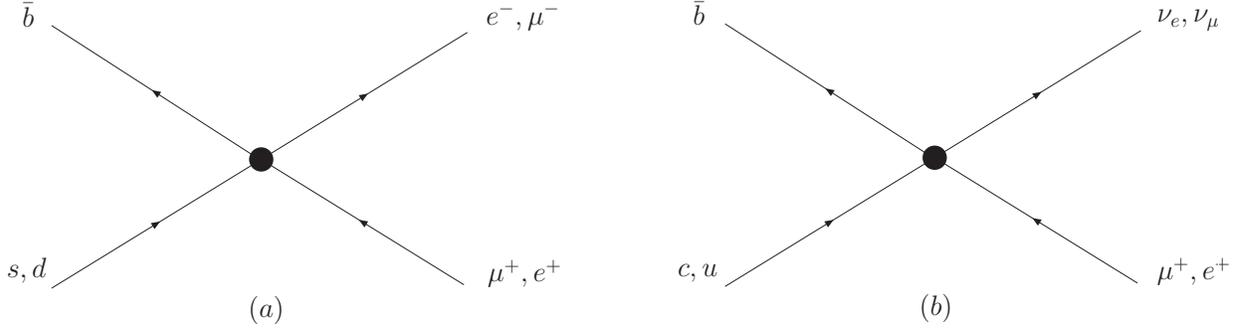,width=6.5in}
\caption{Feynman diagrams for the contributions of the {\it Class Two} operators to (a) $B_0,B_s\rightarrow e^{\pm} \mu^{\mp}$ and (b) $B^+,B^+_c\rightarrow \ell^+_j \nu_k$.}
\label{fig:b2body}
\end{figure}

In this section, we consider the tree-level contributions of the {\it Class Two} operators to the decays of $B$ mesons.  First, we consider the fully leptonic decays, $B^0, B_s \rightarrow e^{\pm} \mu^{\mp}$ and $B^+, B^+_c \rightarrow \ell^+_j \nu_k$.  These are shown in Fig. \ref{fig:b2body}.  It is straightforward to show that operators ${\cal O}_{5,ijk}$ and ${\cal O}_{6,ijk}$ contribute only to $B$ decays with two charged leptons in the final state (shown in Fig. \ref{fig:b2body}(a)), while operators ${\cal O}_{7,ijk}$ and ${\cal O}_{8,ijk}$ contribute only to those in Fig. \ref{fig:b2body}(b) which contain a neutrino in the final state.  
The contributions from the {\it Class Two} operators are found to be (ignoring terms suppressed by $V_{td}$ and $V_{ts}$ and approximating $V_{tb}=1$):
\begin{eqnarray}
\Gamma &=& \frac{1}{64\pi}\frac{|C_{n,ijk}|^2}{\Lambda^4}f^2_{B_{(s)}} m_B m^2_{\mu} |V_{id(s)}|^2\quad\mbox{($n=5$, $6$)} \label{eq:b1}\\
\Gamma &=& \frac{1}{64\pi}\frac{|C_{n,ijk}|^2}{\Lambda^4}f^2_{B_{(c)}}  \frac{m^5_B}{(m_b + m_{u(c)})^2}\quad\mbox{($n=7$)} \label{eq:b2}\\
\Gamma &=& 0 \quad\mbox{($n=8$)} \label{eq:b3}
\end{eqnarray}
where $f_{B_{(s)}}$ is the decay constant taken from \cite{Gray:2005ad} for the $B^0$ or $B_s$, depending on the decay. (Note that we use the convention $f_{\pi^+}\approx 130$ MeV.)  For the charged $B$ decays, we only consider the $B^+$ case, for which we use the same value for the decay constant as for the $B^0$; experimental constraints do not exist for $B^+_c\rightarrow \ell^+_j \nu_k$.  $m_B$ is the $B$ meson mass, $m_b$ and $m_{u(c)}$ are the ($1$S) $b$ quark mass and the $u$ (or $c$) quark mass, respectively.  The factor of $m_{\mu}^2$ in Eq. (\ref{eq:b1}) comes from helicity suppression, which can be traced to operators ${\cal O}_{5,ijk}$ and ${\cal O}_{6,ijk}$ containing lepton fields which are both left-handed (${\cal O}_{5,ijk}$) or both right-handed (${\cal O}_{6,ijk}$).  This helicity suppression does not occur for operator ${\cal O}_{7,ijk}$ which contains lepton fields of opposite chiralities. Operator ${\cal O}_{8,ijk}$ does not contribute to $B$ decay at tree level as the $\sigma_{\mu\nu}$ matrix is antisymmetric, and the only four-vector with which it can be contracted is the $B$ meson four-momentum.
From the current limit on $B^0$ decay \cite{Yao:2006px},
\be
\mbox{Br}(B^0\rightarrow e^{\pm}\mu^{\mp}) \leq 1.7 \times 10^{-7} \quad \mbox{at $90\%$ CL}\nonumber\\
\ee
we derive lower bounds on $|C_{n,ijk}|/\Lambda^2$ for $n=5, 6$:
\begin{eqnarray}
\frac{|C_{5,ujk}|}{\Lambda^2}, \frac{|C_{6,ujk}|}{\Lambda^2} &\leq& \frac{1}{(3.7 \mbox{ TeV})^2}\nonumber\\
\frac{|C_{5,cjk}|}{\Lambda^2}, \frac{|C_{6,cjk}|}{\Lambda^2} &\leq& \frac{1}{(1.8 \mbox{ TeV})^2}.\nonumber
\end{eqnarray}
where the limit on the operators with $i=c$ is weaker than the $i=u$ case due to CKM suppression.  If we had instead considered the decay of the $B_s$ \cite{Yao:2006px},
\be
\mbox{Br}(B_s\rightarrow e^{\pm}\mu^{\mp}) \leq 6.1 \times 10^{-6} \quad \mbox{at $90\%$ CL}\nonumber
\ee
we would have obtained the slightly weaker limit for the $i=c$ case
\be
\frac{|C_{5,cjk}|}{\Lambda^2}, \frac{|C_{6,cjk}|}{\Lambda^2} \leq \frac{1}{(1.6 \mbox{ TeV})^2}.\nonumber
\ee
From the $B^+$ decays
\begin{eqnarray}
\mbox{Br}(B^+\rightarrow e^+ \nu) &\leq& 9.8 \times 10^{-6} \quad \mbox{at $90\%$ CL}\nonumber\\
\mbox{Br}(B^+\rightarrow \mu^+ \nu) &\leq& 1.7 \times 10^{-6} \quad \mbox{at $90\%$ CL},\nonumber
\end{eqnarray}
we obtain
\begin{eqnarray}
\frac{|C_{7,u\mu e}|}{\Lambda^2} &\leq& \frac{1}{(16 \mbox{ TeV})^2}\nonumber\\
\frac{|C_{7,ue\mu}|}{\Lambda^2} &\leq& \frac{1}{(10 \mbox{ TeV})^2}.\nonumber
\end{eqnarray}
Here, we can see that the operators which do not suffer from helicity suppression receive substantially tighter constraints than those which do.

\begin{figure}[h]
\epsfxsize=2in
\epsfig{figure=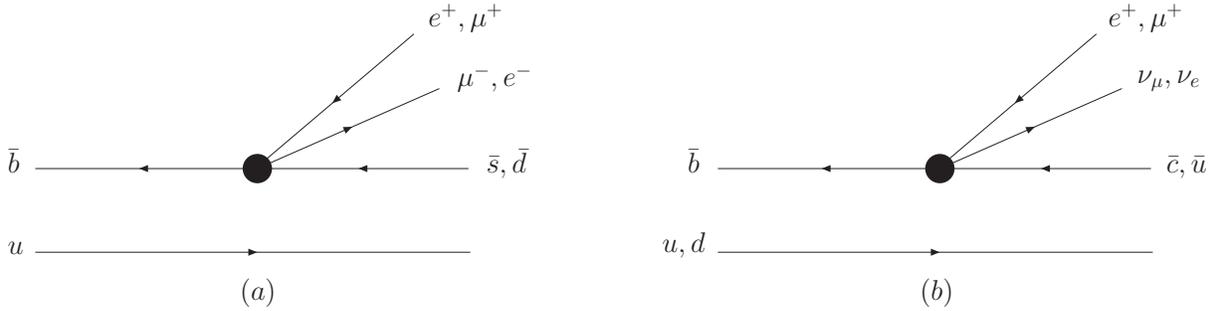,width=6.5in}
\caption{Diagrams for the contributions of the {\it Class Two} operators to the 3-body decays (a) $B^+\rightarrow \pi^+ (K^+) e^{\pm} \mu^{\mp}$ and (b) $B \rightarrow X_{u(c)} \ell^+ \nu$.}
\label{fig:b3body}
\end{figure}

Next, we consider the 3-body decays shown in Fig. \ref{fig:b3body}.  As these decays proceed via a decay of the $b$ quark, the inclusive widths of these processes are the same as in Eq.(\ref{eq:ts}) with $m_t \rightarrow m_b$ (except for the case $B^+\rightarrow X_c e^+ \nu$ below, where we have included effects of the $c$ quark mass).  These processes do not experience the helicity suppression which can occur in the 2-body decays, so we may hope to use these decays to obtain tighter limits on operators ${\cal O}_{5,ijk}$ and ${\cal O}_{6,ijk}$.  Additionally, operator ${\cal O}_{8,ijk}$, which was unconstrained by the 2-body decays, can give a nonzero contribution to the process shown in Fig. \ref{fig:b3body}(b).  

As in the 2-body decays, operators ${\cal O}_{5,ijk}$ and ${\cal O}_{6,ijk}$ will contribute at tree level only to the decays which contain charged leptons in the final state (shown here in Fig. \ref{fig:b3body}(a)).  We compare the exclusive limits \cite{Yao:2006px}
\begin{eqnarray}
\mbox{Br}(B^+\rightarrow\pi^+e^+\mu^-) &\leq& 6.4 \times 10^{-3}\quad\mbox{at $90\%$ CL}\nonumber\\
\mbox{Br}(B^+\rightarrow\pi^+e^-\mu^+) &\leq& 6.4 \times 10^{-3}\quad\mbox{at $90\%$ CL}\nonumber\\
\mbox{Br}(B^+\rightarrow K^+e^{\pm}\mu^{\mp}) &\leq& 9.1 \times 10^{-8}\quad\mbox{at $90\%$ CL}\nonumber
\end{eqnarray} 
with those of the SM process $\Gamma(B\rightarrow \pi \ell^+ \nu)$ and relate them to our inclusive calculations by estimating that 
\be
\frac{\Gamma((B^+\rightarrow\pi^+e^{\pm}\mu^{\mp})}{\Gamma(B^+\rightarrow X_d e^{\pm}\mu^{\mp})} \approx \frac{\Gamma(B\rightarrow \pi \ell^+ \nu)}{\Gamma(B\rightarrow X_u \ell^+ \nu)}\approx \frac{\Gamma((B^+\rightarrow K^+e^{\pm}\mu^{\mp})}{\Gamma(B^+\rightarrow X_s e^{\pm}\mu^{\mp})} .\nonumber
\ee
From the decay $B^+\rightarrow K^+e^{\pm}\mu^{\mp}$, we obtain
\begin{eqnarray}
\frac{|C_{5,ujk}|}{\Lambda^2}, \frac{|C_{6,ujk}|}{\Lambda^2} &\leq& \frac{1}{(8 \mbox{ TeV})^2}\nonumber\\
\frac{|C_{5,cjk}|}{\Lambda^2}, \frac{|C_{6,cjk}|}{\Lambda^2} &\leq& \frac{1}{(16 \mbox{ TeV})^2}.\nonumber
\end{eqnarray}
Note that the limit on the $i=u$ case from $B^+\rightarrow K^+e^{\pm}\mu^{\mp}$ is quite strong, despite being CKM-suppressed.  If we instead look at the non-CKM-suppressed processes with $\pi^+$'s in the final state, we would obtain limits $|C_{5(6),ujk}|/\Lambda^2 < 1/(1\mbox{ TeV})^2$.

Operators ${\cal O}_{7,ijk}$ and ${\cal O}_{8,ijk}$ can contribute to the processes $B\rightarrow X_{u(c)} \ell^+ \nu$ through the diagram in Fig \ref{fig:b3body}(b).  Although these processes violate lepton flavor, they are experimentally indistinguishable from SM processes.  Here, to obtain a rough estimate on the possible size of the operator coefficients, we take $2\times$ the experimental error on these processes as an approximate size of the contribution of NP.  Using the experimental measurements \cite{Yao:2006px},
\begin{eqnarray}
\mbox{Br}(B\rightarrow X_u \ell^+ \nu) &=& 2.33 \pm .22 \times 10^{-3}\nonumber\\
\mbox{Br}(B^+\rightarrow X_c e^+ \nu) &=&  10.8 \pm 0.4\%\nonumber
\end{eqnarray}
we obtain
\begin{eqnarray}
\frac{|C_{7,ujk}|}{\Lambda^2} &\leq& \frac{1}{(3 \mbox{ TeV})^2}\nonumber\\
\frac{|C_{7,ce\mu}|}{\Lambda^2} &\leq& \frac{1}{(1 \mbox{ TeV})^2}\nonumber\\
\frac{|C_{8,ujk}|}{\Lambda^2} &\leq& \frac{1}{(7 \mbox{ TeV})^2}\nonumber\\
\frac{|C_{8,ce\mu}|}{\Lambda^2} &\leq& \frac{1}{(3 \mbox{ TeV})^2}.\nonumber
\end{eqnarray}
We also take the limits given above for the $i, j, k=c, e, \mu$ case as approximately valid for the $i, j, k=c, \mu, e$ case, as measurements of the exclusive process $B^+\rightarrow \bar{D}^0 \ell^+ \nu$ \cite{Yao:2006px} would yield only slightly weaker results. 

\section{Kaon Decays}
\label{sec:ks}
Now, we consider constraints on our operators which can be obtained from $K_L$ decays via both tree-level and one-loop diagrams.  Although the contributions of these operators to $K_L$ decay will be suppressed by small CKM factors, the decay width $\Gamma(K_L\rightarrow e^{\pm} \mu^{\mp})$ is constrained to be $\sim 9$ orders of magnitude smaller than the analogous decay width of the $B$ mesons. Thus, it is worthwhile to consider both tree-level and one-loop diagrams contributing to $K_L$ decay.  Here, unlike the purely tree-level $B$ decay case, we may hope to derive limits on the {\it Class One} operators, in addition to those in {\it Class Two}. 

First, we consider tree-level contributions to $K_L\rightarrow e^{\pm} \mu^{\mp}$.  As in the $B$ case, only {\it Class Two} operators can contribute, as the {\it Class One} operators contain right-handed top quarks.  ${\cal O}_{7,8jk}$ and ${\cal O}_{8,ijk}$ cannot contribute at tree-level to this decay, as $K_L$ contains two down-type quarks, and these operators contain a right-handed $u_R$.  ${\cal O}_{5,ijk}$ and ${\cal O}_{6,ijk}$, however, can contribute at tree level to this decay.  Here, we will specifically discuss ${\cal O}_{5,ijk}$.  The results for ${\cal O}_{6,ijk}$ (at both tree and one-loop level) are identical to those of ${\cal O}_{5,ijk}$.

To simplify the discussion, we introduce another operator similar to ${\cal O}_{5,ijk}$ but which contains left-handed $d$ and $s$ quarks
\be
{\cal O}_{5,Kjk} = \bar{d}_L \gamma^{\mu} s_L \bar{L}^j_L \gamma_{\mu} L^k_L \nonumber\\
\ee
where the $d$ and $s$ quarks in this operator are in the mass eigenstate basis.  This operator will contribute to $K_L$ decay as
\be
\Gamma(K_L\rightarrow e^{\pm} \mu^{\mp}) = \frac{|C_{5,Kjk}|^2}{\Lambda^4} \frac{1}{64\pi m_K^3}(m_K^2-m_{\mu}^2)^2 m_{\mu}^2 F_K^2
\ee
where the factor of $m^2_{\mu}$ is due to helicity suppression.  The experimental limit \cite{Yao:2006px}
\be
\mbox{Br}(K_L \rightarrow e^{\pm} \mu^{\mp}) \leq 4.7\times 10^{-12},
\ee
yields the very tight constraint
\be
\frac{|C_{5,Kjk}|^2}{\Lambda^4} < \frac{1}{(320 \mbox{ TeV})^4}
\ee

The quark doublets $\bar{Q}^i_L$ and $T_L$ in ${\cal O}_{5,ijk}$ have, as lower components, $V^*_{id} d + V^*_{is} s + V^*_{ib} b$, and $V_{td} d + V_{ts} s + V_{tb} b$, respectively.  Thus, ${\cal O}_{5,ijk}$ contains, at tree level, a term 
\be
\label{eq:k1}
V^*_{id} V_{ts} {\cal O}_{5,Kjk} 
\ee
and a term
\be
\label{eq:k2}
V^*_{is} V_{td} {\cal O}^{\dagger}_{5,Kkj}. 
\ee
Both of these terms contribute to the same final state and, thus, will interfere,
\be
\frac{|C_{5,ijk}|^2}{\Lambda^4} |V^*_{id} V_{ts} + V^*_{is} V_{td}|^2  < \frac{1}{(320 \mbox{ TeV})^4},
\ee
and, thus, we obtain the very tight constraints
\begin{eqnarray}
\frac{|C_{5,ujk}|^2}{\Lambda^4} &<& \frac{1}{(60\mbox{ TeV})^4}\nonumber\\
\frac{|C_{5,cjk}|^2}{\Lambda^4} &<& \frac{1}{(40\mbox{ TeV})^4}\nonumber.
\end{eqnarray}

\begin{figure}[h]
\epsfxsize=2in
\epsfig{figure=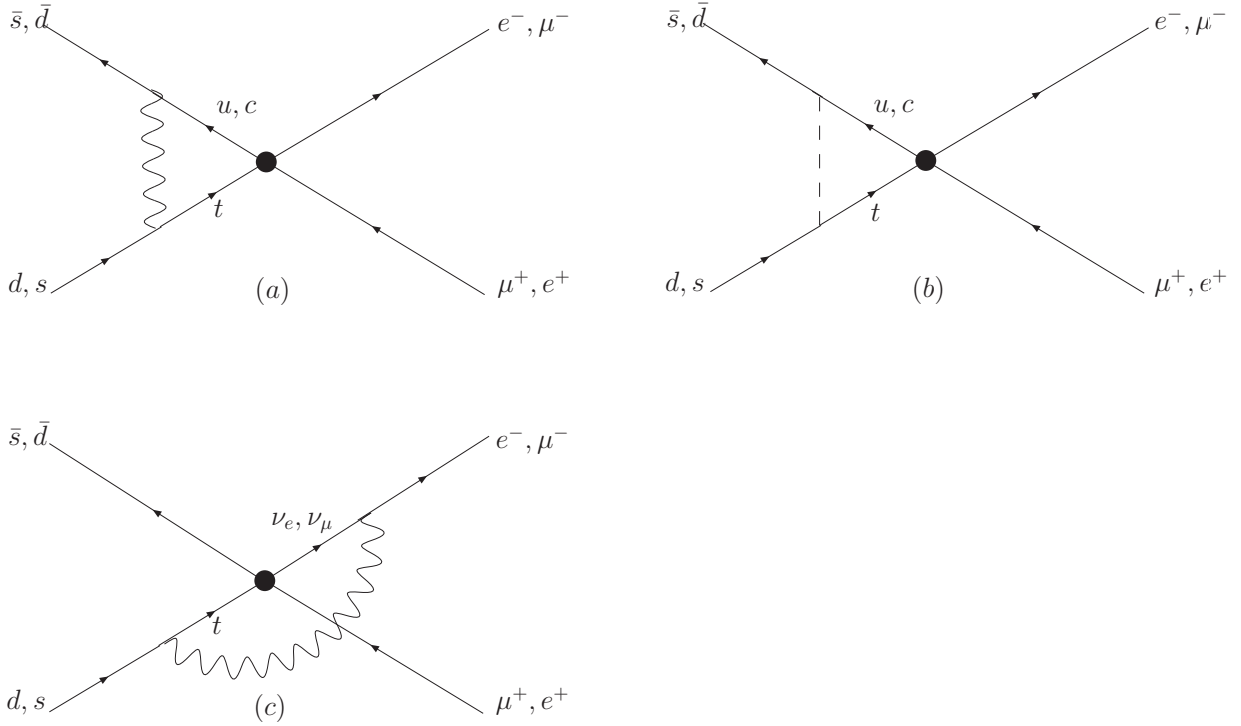,width=6.5in}
\caption{Diagrams contributing to $K_L \rightarrow e^{\pm} \mu^{\mp}$.}
\label{fig:ks}
\end{figure}

We now turn to one-loop diagrams.  Representative diagrams which can contribute to $K_L\rightarrow e^{\pm} \mu^{\mp}$ are shown in Fig. \ref{fig:ks}.  
In general, several factors contribute to the relative size of these diagrams for a given operator.  First, let us consider the diagram in Fig. \ref{fig:ks}(a).  If the operator insertion in (a) contains lepton fields of the same chirality (as in operators ${\cal O}_{1,ijk}$, ${\cal O}_{2,ijk}$, ${\cal O}_{5,ijk}$, and ${\cal O}_{6,ijk}$), this diagram, when inserted into $K_L$ decay, will give an amplitude which is helicity-suppressed, similarly to the 2-body decays discussed above.  Operators ${\cal O}_{3,ijk}$, ${\cal O}_{4,ijk}$, ${\cal O}_{7,ijk}$, and ${\cal O}_{8,ijk}$, on the other hand, will suffer no such suppression when inserted into this diagram.  Secondly, as the SM vertices in (a) are purely left-handed, the insertion of those operators with right-handed quark fields into (a) will induce factors of $m_{u(c)}$ and/or $m_t$ due to the quark spin flip.  On the other hand, the diagram in Fig \ref{fig:ks}(b) will have a different dependence on the quark masses, as the SM vertices in this diagram are not purely left-handed and the Yukawa couplings themselves give factors of $m_t$ or the light-quark masses; however, the diagram can still suffer from helicity suppression, depending upon the operator inserted.  Finally, the diagram in (c) is relevant only for operators ${\cal O}_{3,ijk}$ and ${\cal O}_{4,ijk}$.  As these operators contain lepton fields of opposite chirality, their contributions to $K$ decay via this diagram will never be helicity-suppressed, and, as each of these operators contains a $t_R$ field, the decay amplitudes given by these diagrams will contain a factor of $m_t$.  Also, we note that all of these diagrams will contain flavor-dependent CKM factors $V_{ij}$.  

We will discuss two specific cases of operator insertions into these diagrams.  First, we will discuss the insertion of operator ${\cal O}_{5,ijk}$ into diagram \ref{fig:ks}(a).  In this case, both quark fields in the operator are left-handed, so there are no factors of quark mass which come from forcing a fermion inside the loop to flip chirality between the SM and NP vertices, and the diagram is logarithmically divergent.  We estimate the mixing of ${\cal O}_{5,ijk}$ into ${\cal O}_{5,Kjk}$ via diagram \ref{fig:ks}(a) and obtain
\be
C_{5,Kjk}(v) \approx C_{5,ujk}(\Lambda) \frac{\alpha}{2 \sin^2 \theta_W} (V_{ud}^* V_{ts}) \frac{1}{(4\pi)} \ln{\frac{v^2}{\Lambda^2}}
\ee
where the factor of $\ln{v^2/\Lambda^2}$ comes from running of the operator coefficient from the scale $\Lambda$ to the weak scale $v$.  (Here, we have ignored the running of this coupling from $v$ down to the $K$ mass due to photon and gluon loops.)  Comparing with Eqs.(\ref{eq:k1}) and (\ref{eq:k2}), we can see that this will not compete with the tree-level result.

Next, we consider the case of the insertion of operator ${\cal O}_{4,ijk}$ into diagram (c). (Although operator ${\cal O}_{3,ijk}$ is similar, it turns out to be suppressed relative to operator ${\cal O}_{4,ijk}$ by a numerical factor.)  This operator will not suffer helicity suppression when it is inserted into diagram (c).  As the operator contains a $t_R$ field, the top quark in the diagram must flip helicity; this gives the amplitude for this diagram a factor of $m_t$ and renders the momentum integral convergent.  The amplitude contains factors of the quark and lepton momentum, which we take of order $m_K$.  We obtain, from this diagram (as well as a similar contribution from diagram (a)), the order-of-magnitude limit
\be
\frac{C_{4,ujk}}{\Lambda^2}\lesssim\frac{1}{(1\mbox{ TeV})^2}
\ee
with a weaker limit for $\frac{|C_{4,cjk}|}{\Lambda^2}$.

For all other operators, we found limits on $\Lambda$ from one-loop diagrams which were weaker than $\sim 1$ TeV.

As the constraints obtained for operators ${\cal O}_{5,ijk}$ and ${\cal O}_{6,ijk}$ from $K_L\rightarrow e^{\pm}\mu^{\mp}$ were helicity-suppressed, one might also consider attempting to constrain these operators via their contributions to 3-body $K_L$ decays, for which helicity suppression is not an issue.  We estimated the contribution of operator ${\cal O}_{5,ijk}$ to $K^-\rightarrow \pi^- e^{\pm} \mu^{\mp}$ and found a weaker result than that from the 2-body decay.  As we do not expect a comparably large improvement in the constraints on the other operators by considering 3-body decays, we do not pursue the 3-body $K_L$ decay results further.

\section{$\mu\rightarrow e$ Conversion in Nuclei and $\mu\rightarrow e \gamma$}
\label{sec:muegamma}
Finally, we briefly consider the contributions of our operators to the processes $\mu\rightarrow e$ conversion in nuclei and $\mu\rightarrow e \gamma$.   $\mu\rightarrow e$ conversion in nuclei can take place via diagrams identical to those in $K_L$ decay, but with both quarks $d$ or $s$ flavor.  ($\mu\rightarrow e$ conversion can also occur via diagrams containing two external $u$ quarks; however, as our operators all contain top quarks, diagrams with two external $u$ quarks can only occur at loop level with factors of $V_{tl}V^*_{ul}$ with $l$ summed over $d, s, b$; these amplitudes are highly suppressed and for our purposes can be neglected.)  As in $K_L$ decay, operators ${\cal O}_{5,ijk}$ and ${\cal O}_{6,ijk}$ contribute at tree level, whereas all other operators contribute first at one-loop level.  Generally, the constraints on $C_{n,ijk}/\Lambda^2$ derived from the current limits on $\mu\rightarrow e$ conversion \cite{Dohmen:1993mp} are less than or comparable to those derived from $K_L$ decay; in particular, for the loop-suppressed operators ($n\ne 5, 6$), limits on $C_{n,ijk}/\Lambda^2$ were weaker than $\lesssim 1/(1 \mbox{ TeV})^2$.  This would, however, not be the case given results from the possible future mu2e experiment \cite{Carey:2007zz}, which intends to improve on the limits of \cite{Dohmen:1993mp} by more than $4$ orders of magnitude; this would correspond to limits on $C_{n,ijk}/\Lambda^2$ of a few hundred TeV for $n=5,6$ and limits for some of the loop-suppressed operators as high as $\sim 10$ TeV .

\begin{figure}[h]
\epsfxsize=2in
\epsfig{figure=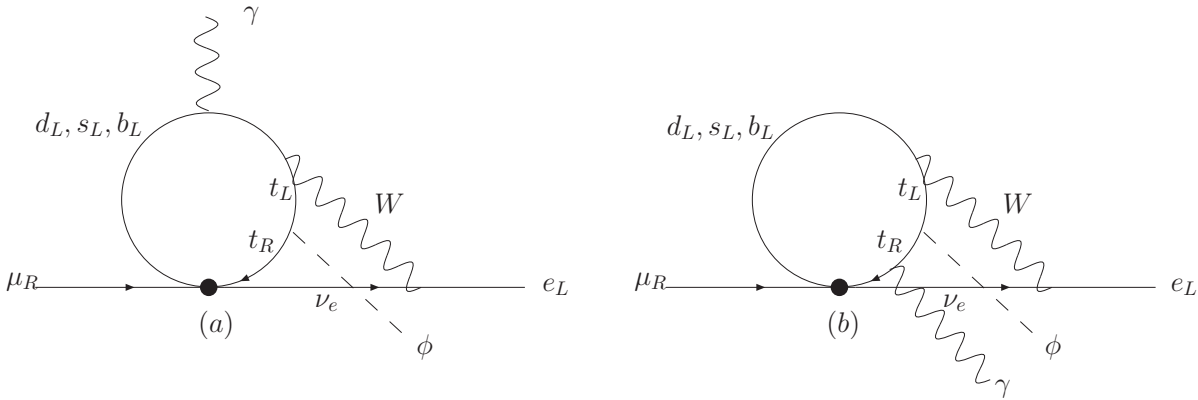,width=6.5in}
\caption{Representative two-loop contributions of operators ${\cal O}_{3,ijk}$ and ${\cal O}_{4,ijk}$ to $\mu\rightarrow e \gamma$.  A Higgs insertion has been shown explicitly to represent the top quark chirality flip.}
\label{fig:mueg}
\end{figure}

Contributions of our operators to $\mu\rightarrow e \gamma$, on the other hand, arise first at two-loop level, due to the quark fields being of different flavor; representative diagrams are shown in Fig. \ref{fig:mueg}.   We neglect diagrams which contain Yukawa couplings to any fermion except the top and bottom quarks.  Note that these diagrams require that the external electron and muon fields have opposite chiralities; diagrams which instead give mixing of our operators into an operator of the form $\bar{e}{\not\!\! D}\mu$ can be absorbed by a redefinition of the fields in the SM kinetic terms.  Similarly, terms of the form $\bar{e}\mu$ just contribute to the lepton mass matrix.  Instead, we must calculate the contribution of our operators which takes the form of the effective transition magnetic moment operator
\be
{\cal O}_{F,jk}=e\frac{v}{\sqrt{2}}\bar{\ell}^j_L \sigma^{\mu\nu} \ell^{k}_R F_{\mu\nu}\nonumber\\
\ee
where $j,k= e,\mu$ or $\mu,e$, $F_{\mu\nu}$ is the QED field strength tensor, and $v$ is the Higgs vacuum expectation value.  We have included the factor of $v/\sqrt{2}$ in ${\cal O}_{F,jk}$ as this operator follows from a linear combination of
\begin{eqnarray}
{\cal O}_{B,jk}&=&g' \bar{L}^j_L \phi \sigma^{\mu\nu} \ell^{k}_R B_{\mu\nu}\nonumber\\
{\cal O}_{W,jk}&=&g \bar{L}^j_L \tau^a \phi \sigma^{\mu\nu} \ell^{k}_R W_{\mu\nu}^a\nonumber
\end{eqnarray}
wherein, after electroweak symmetry breaking, $\phi\rightarrow \left( \begin{array}{c} 0 \\ v/\sqrt{2} \end{array} \right)$.  ${\cal O}_{F,jk}$ will, in turn, contribute to $\mu\rightarrow e \gamma$:
\begin{equation}
\Gamma(\mu\rightarrow e \gamma) = \frac{|C_{F,jk}|^2}{\Lambda^4}\frac{ e^2 v^2 m_{\mu}^3}{8\pi}.\nonumber
\end{equation}
Given the current experimental constraint,
\begin{equation}
Br(\mu\rightarrow e \gamma) < 1.2 \times 10^{-11} \mbox{at $90\%$ CL\cite{Brooks:1999pu}},\nonumber
\end{equation}
the coefficient of ${\cal O}_{F,jk}$ is constrained to be
\begin{equation}
\frac{|C_{F,jk}|}{\Lambda^2} < \frac{1}{(16,000\mbox{ TeV})^2}.
\end{equation}
We will see, however, that, although this limit is quite strong, the contributions from our operators are very highly suppressed.  

Neglecting terms suppressed by Yukawa couplings other than those of the top and bottom quarks, we find that only the operators ${\cal O}_{3,ijk}$ and ${\cal O}_{4,ijk}$ contribute to the process $\mu \rightarrow e \gamma$.  This can be seen as follows.  First, we require operators which contain lepton fields of opposite chirality, which limits us to the scalar and tensor operators ${\cal O}_{3,ijk}$, ${\cal O}_{4,ijk}$, ${\cal O}_{7,ijk}$, and ${\cal O}_{8,ijk}$.  All of these operators contain one right-handed up-type quark field and one left-handed quark field.  The two quark fields in the operator must be placed in a loop as shown in example diagrams in Fig. \ref{fig:mueg}.  As the quarks are from different families, this loop must be accompanied by an internal line which is either an $W$ boson or the charged component of the Higgs field.  If the diagram contains a $W$, then the up-type right-handed quark must flip its chirality, which gives a factor of the quark mass.  If the diagram instead contains an internal Higgs line, then the coupling of the Higgs to the right-handed up-type quark contains a factor of the same quark mass.  If we neglect the masses of the first and second generation up-type quarks, our diagrams must contain a right-handed top quark, $t_R$, which is contained only in ${\cal O}_{3,ijk}$ and ${\cal O}_{4,ijk}$. 

We give some examples of diagrams which contribute to $\mu \rightarrow e \gamma$ in Fig. \ref{fig:mueg}.  Although the diagrams in Fig. \ref{fig:mueg} are superficially logarithmically divergent, it is important to notice that the amplitudes from these diagrams contain factors of $V_{tl}V^*_{il}$, where $i=u,c$ and $l=d,s,b$.  If there is no dependence on the mass of the internal quark of flavor $l$, then, due to GIM unitarity, this expression vanishes when summed over $l$.  As the logarithmic divergence is independent of $l$, the contributions of these diagrams is rendered finite.  Thus, as we expect some suppression by the large $W$ and $t$ masses, we expect $\frac{|C_{F,jk}|}{\Lambda^2}$ to go as
\begin{equation}
\label{eq:muegoom}
\frac{|C_{F,jk}|}{\Lambda^2} \approx \frac{|C_{n,ijk}|}{\Lambda^2} \frac{g^2}{(4\pi)^4} \frac{m_b^2}{m_w^2} |V_{tb}V^*_{ib}|.
\end{equation}
As the largest value the CKM factors can take is $|V_{tb}V^*_{cb}|\approx 4\times 10^{-2}$ (for $i=c$), this suppresses the physics reach by a factor of approximately $2\times 10^{4}$, which yields the order-of-magnitude estimate 
\begin{equation}
\frac{|C_{n,cjk}|}{\Lambda^2} \lesssim \frac{1}{(1\mbox{ TeV})^2}\nonumber
\end{equation}
with a weaker limit for the $i=u$ case.  In the case of operator ${\cal O}_{4,ijk}$, there also exist diagrams which have similar suppression as that given in Eq. (\ref{eq:muegoom}), but which are indeed logarithmically divergent.  We expect that these contributions to  $\frac{|C_{F,jk}|}{\Lambda^2}$ will thus be enhanced by a factor $\sim \ln{v^2/\Lambda^2}$.  For the $i=c$ case, the limit on $\Lambda$ from $\mu\rightarrow e \gamma$ can thus be comparable to, but not substantially more than, that expected from top decay, of order a few TeV.

\section{Conclusions}
\label{sec:conc}
We consolidate our results in Table \ref{table:results}, which lists our operators, the expected lower bounds on $\Lambda$ from $t$ decays with $C_{n,ijk}$ set equal to unity, and the current lower bounds from $B$ decays and $K$ decays.  We note that, of the $32$ possible contributing operators ($8$ operator structures $\times$ $4$ flavor combinations), only $12$ are currently constrained significantly more tightly than we would expect from $t$ decay.  In particular, the {\it Class One} operators are largely unconstrained by current measurements.  Thus, it is possible that operators which could produce observable, flavor-violating results in top decays at LHC have so far evaded experimental constraints. 
\begin{table}
\caption{Lower bounds on $\Lambda$, in TeV, with $C_{n,ijk}=1$.  In the first column, we list the operators, specifying flavor indices where relevant.  The second column lists the expected limits from top decay, the third and fourth columns list the results from the 2-body and 3-body $B$ decays, respectively, and the fifth column gives the results from $K_L$ decay.  Dashes indicate that no constraint exists from a given observable, or that the constraint is too weak to be interesting ($\lesssim 1$ TeV).}  
\label{table:results}
\begin{tabular}{|l|c|c|c|c|}
\hline
Operator & $t\rightarrow u(c)e^{\pm}\mu^{\mp}$ & $B$, 2-body & $B$, 3-body & $K_L$ \\
\hline
${\cal O}_{1,ijk}=\bar{u}_R^i \gamma^{\mu} t_R \bar{L}^j_L \gamma_{\mu} L^k_L $&2.1&-&-&-\\
\hline
${\cal O}_{2,ijk}=\bar{u}_R^i \gamma^{\mu} t_R \bar{l}^j_R \gamma_{\mu} l^k_R$&2.1&-&-&-\\
\hline
${\cal O}_{3,ijk}=\epsilon^{ab} \bar{Q}_{La}^i  t_R \bar{L}^j_{Lb}  l^k_R$&&&&\\
\quad$i=u$&1.5&-&-&-\\
\quad$i=c$&1.5&-&-&-\\
\hline
${\cal O}_{4,ijk}=\epsilon^{ab} \bar{Q}_{La}^i \sigma^{\mu\nu} t_R \bar{L}^j_{Lb} \sigma_{\mu\nu} l^k_R$&&&&\\
\quad$i=u$&4.0&-&-&1\\
\quad$i=c$&4.0&-&-&-\\
\hline
\hline
${\cal O}_{5,ijk}=\bar{Q}^i_L \gamma^{\mu} T_L \bar{L}^j_L \gamma_{\mu} L^k_L$&&&&\\
\quad $i=u$&2.1&3.7&8&60\\
\quad $i=c$&2.1&1.8&16&40\\
\hline
${\cal O}_{6,ijk}=\bar{Q}^i_L \gamma^{\mu} T_L \bar{l}^j_R \gamma_{\mu} l^k_R$&&&&\\
\quad $i=u$&2.1&3.7&8&60 \\
\quad $i=c$&2.1&1.8&16&40\\
\hline
${\cal O}_{7,ijk}=\epsilon^{ab} \bar{u}_R^i T_{La} \bar{l_R}^j  L^k_{Lb}$&&&&\\
\quad $i=u, j=e, k=\mu$&1.5&10&3&-\\
\quad $i=u, j=\mu, k=e$&1.5&16&3&-\\
\quad $i=c$&1.5&-&1&-\\
\hline
${\cal O}_{8,ijk}=\epsilon^{ab} \bar{u}_R^i \sigma^{\mu\nu} T_{La} \bar{l_R}^j \sigma_{\mu\nu} L^k_{Lb}$&&&&\\
\quad $i=u$&4.0&-&7&-\\
\quad $i=c$&4.0&-&3&-\\
\hline
\end{tabular}
\end{table}

One issue which we have not addressed here is the backgrounds relevant to the process $t\rightarrow (c) e^{\pm} \mu^{\mp}$ at LHC.  Here, we are interested in $t\bar{t}$ production where one of the top quarks exhibits this flavor-violating decay, and the other decays as in the SM, $t\rightarrow b W^+$.  The experimental signature of this decay is a light jet, and an isolated muon and an isolated electron of opposite charge, the three of which give a combined invariant mass consistent with that of a top quark, along with a $b$ jet and the decay products of a $W$, which can be either leptonic (a charged lepton plus missing energy) or hadronic (two light jets).  Additionally, in the case where the $W$ decays hadronically, the decay products of the top quark pair should contain very little missing $p_T$.

A possible SM background to this process is $t\bar{t}$ production where the $W$'s from both top decays decay leptonically; this could give a final state with two isolated, opposite-sign leptons of different flavor and 2 $b$-jets.  However, this process will give a final state with substantial missing $p_T$, due to the two missing neutrinos from the $W$ decays, and the lepton pair plus one of the $b$-jets will not reconstruct an invariant mass of $\approx m_t$. 

A study by ATLAS \cite{Chikovani:2002cg} (see also \cite{Carvalho:2007yi}) of a somewhat similar NP signature, $t\rightarrow c Z$, followed by the SM decay $Z\rightarrow \ell^+ \ell^-$ (with both leptons of the same flavor), found the main backgrounds to their process to be $Z+jets$, $WZ$ and $t\bar{t}$.  Their analysis concluded that branching fractions of the $t\rightarrow c Z$ could be measured at the $10^{-4}$ level.  However, as their analysis required that the lepton pair be consistent with an SM decay of the $Z$, {\it i.e.} that there be leptons of the same flavor with a combined invariant mass compatible with that of the $Z$, their results are not exactly analogous with those one would expect for the process $t\rightarrow u(c) e^{\pm}\mu^{\mp}$, where, in particular, the requirement of two leptons of different flavor would be expected to significantly reduce backgrounds containing a $Z$ decay.  Although we take an estimate of an observable branching fraction of $t\rightarrow u(c) e^{\pm}\mu^{\mp}$ to be $10^{-7}$, further study would be necessary to determine the possible precision of this measurement.  We note that, as the branching fraction for $t\rightarrow u(c) e^{\pm}\mu^{\mp}$ goes as $\sim 1/\Lambda^4$, if the observable branching fraction turned out to be, for example, $10^{-6}$, this would reduce the lower bound on $\Lambda$ by a factor of $\sim 1.8$. 

Also, although we have done a model-independent analysis, one may ask about the constraints this analysis places on specific models.  Although lepton-flavor-violating decays of the top quark have been explored in some models (see Refs. \cite{Iltan:2002re,Iltan:2004js,DiazCruz:2004sp,Yue:2004jc}), decays containing $e^{\pm}\mu^{\mp}$ have, so far, not been shown to be substantial.  However, as we are doing a model-independent analysis of unknown physics beyond the SM, and as these decays are not experimentally ruled out, we do not take this as an indication that these decays are not experimentally interesting. 

Also, one may also consider exotic, but less exotic, processes which contain a top and a light quark yet which conserve lepton flavor; one could guess that NP models which give $t\rightarrow (c) e^{\pm} \mu^{\mp}$ would also give analogous lepton-flavor-conserving processes, and set limits on the scale of NP accordingly.  The L3 collaboration \cite{Achard:2002vv} has studied $e^+e^-\rightarrow t \bar{u} (\bar{c})$ and found lower bounds on $\Lambda$ for scalar, vector, and tensor operators.  Their strongest limit is for the tensor case, with a final state containing a $c$ quark; for this case, they have $\Lambda\gtrsim 1.24$ TeV, for $m_t=170$ GeV.  For most of $B$ and $K_L$ decays studied here, the analogous lepton-flavor-conserving processes are not substantially more constraining that those which we have considered.  The largest improvement would come from considering the decay $B_s \rightarrow \mu^+\mu^-$, whose branching fraction of $<1.5\times 10^{-7}\mbox{($90\%$ CL)}$ \cite{Yao:2006px}, is approximately a factor of $\sim 40$ more strongly constrained than the process containing $e^{\pm}\mu^{\mp}$ (which receives contributions from operators ${\cal O}_{5,ijk}$ and ${\cal O}_{6,ijk}$).  As the branching fraction is proportional to $\frac{1}{\Lambda^4}$, this process would give a limit on $\Lambda$ approximately $\sim 3$ times stronger than the limit from the lepton-flavor-violating $B_s$ decay.

Finally, we note that our calculations for $\Gamma(t\rightarrow u(c) e^{\pm} \mu^{\mp})$ would have to be altered if it turns out that the scale of new physics is not much greater than $m_t$ (which could happen if, for example, new particles which could mediate $t\rightarrow u(c) e^{\pm} \mu^{\mp}$ can be produced at energies accessible at LHC).  However, in that event, one would hope that there would be other signatures of new physics awaiting discovery in the LHC data.

\section{Acknowledgements}
The authors would like to thank Shrihari Gopalakrishna, Frank Paige and Kyle Cranmer for their helpful comments, and Peter Yamin and Andre de Gouvea for providing information on the mu2e experiment and $\mu\rightarrow e$ conversion.  This work is supported in part by US DOE contract \# DE-AC02-98CH10886.
  
\bibliographystyle{h-physrev}

\end{document}